\documentstyle[12pt,twoside,fleqn,espcrc1]{article}

\newcommand{\AmS}{{\protect\the\textfont2
  A\kern-.1667em\lower.5ex\hbox{M}\kern-.125emS}}

\input{epsf}
\title{Production of Stable and Unstable Light Nuclei and Hyperfragments
       in 11.5 A GeV/c Au+Pb collisions}

\author{L. Evan Finch, Yale University, for the E864 Collaboration}

\begin{document}
\maketitle

\begin{abstract}
We present measurements of the production of stable light nuclei for
mass number $A \leq 7$, of strongly decaying states $^{5}He$ 
and $^{5}Li$ and of the the hypernucleus $^{3}_{\Lambda}H$.  
We also examine trends in the production of these multibaryon
states as a  function of kinematic variables and 
properties of the states including strangeness content. 
\end{abstract}

\vspace{-0.05in}
\section{INTRODUCTION}

Experiment E864 at Brookhaven has a quite comprehensive set of
measurements which addresses the topic of coalescence of multibaryon
states in heavy-ion collisions at AGS energies.  These 
include measurements of stable light nuclei from mass number $A$=1 (including 
the first measurement of neutrons in an AGS Au+Pb or similar system) up to 
$A$=7.  In addition, we have measured production of the
strongly decaying states $^{5}He$ and $^{5}Li$ and the hypernucleus
$^{3}_{\Lambda}H$.  

With this set of measurements we can systematically
examine the dependences of production of coalesced states on 
kinematic variables and collision centrality as well as on the
mass number, spin, isospin, and strangeness content of the 
state.

\vspace{-0.05in}
\section{KINEMATIC AND CENTRALITY DEPENDENCES}

Only for the lighter states ($A \leq 3$) do we currently have
sufficient statistics to study the
kinematic and centrality dependences of production.   

\begin{figure}[htb]
\begin{minipage}[b]{.46\linewidth}
\centering\epsfxsize=3in \epsfysize=3in \leavevmode \epsfbox[10 154 534 654]{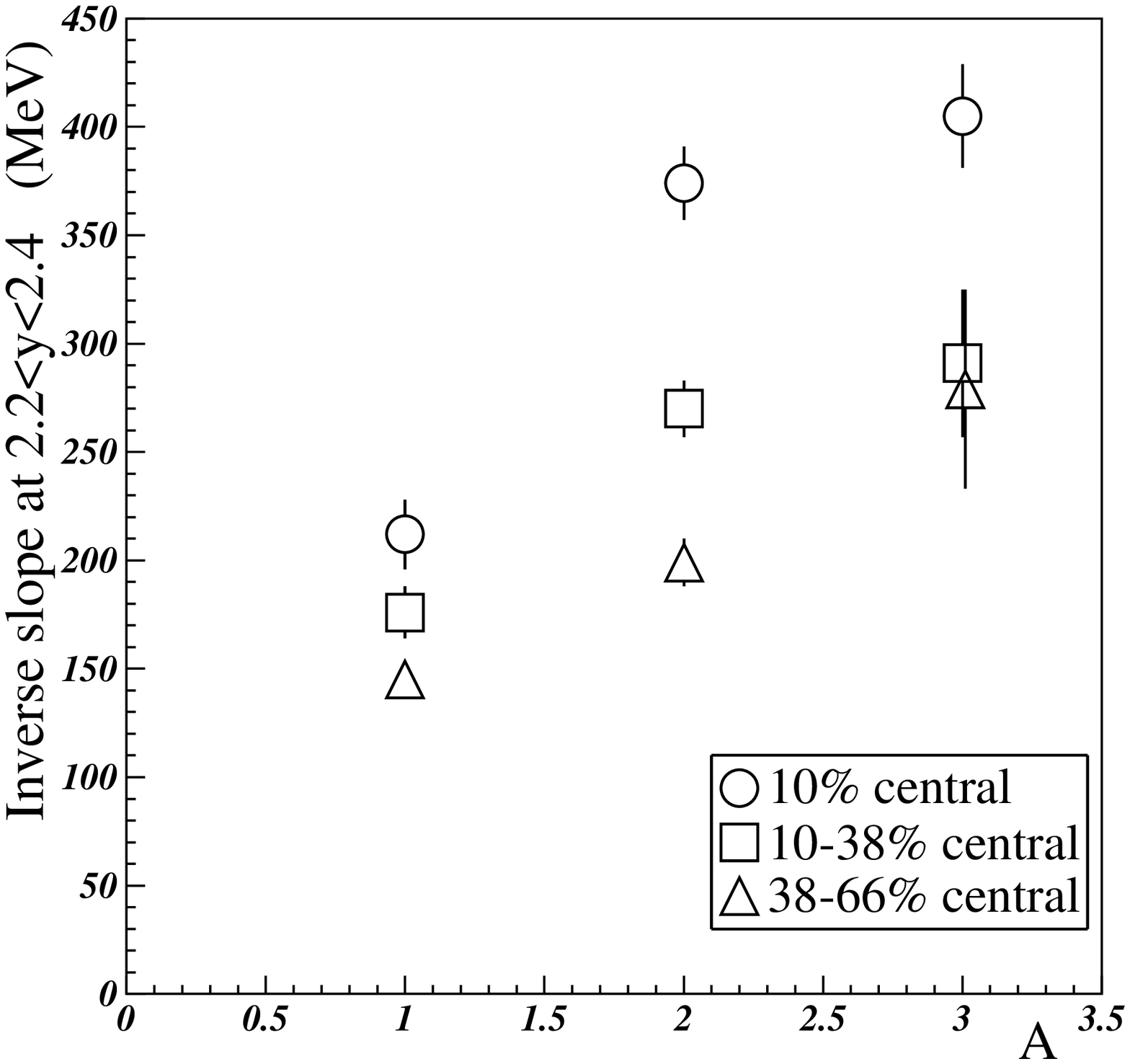}
\vspace{-0.45in}
\caption{Inverse slope parameters as a function of mass number for three different event centralities in the 
rapidity bin 2.2$<y<$2.4.}
\label{fig:temp_vs_mass}
\end{minipage}
\hspace{\fill}
\begin{minipage}[b]{.46\linewidth}
\centering\epsfxsize=3in \epsfysize=3in \leavevmode \epsfbox[10 154 534 654]{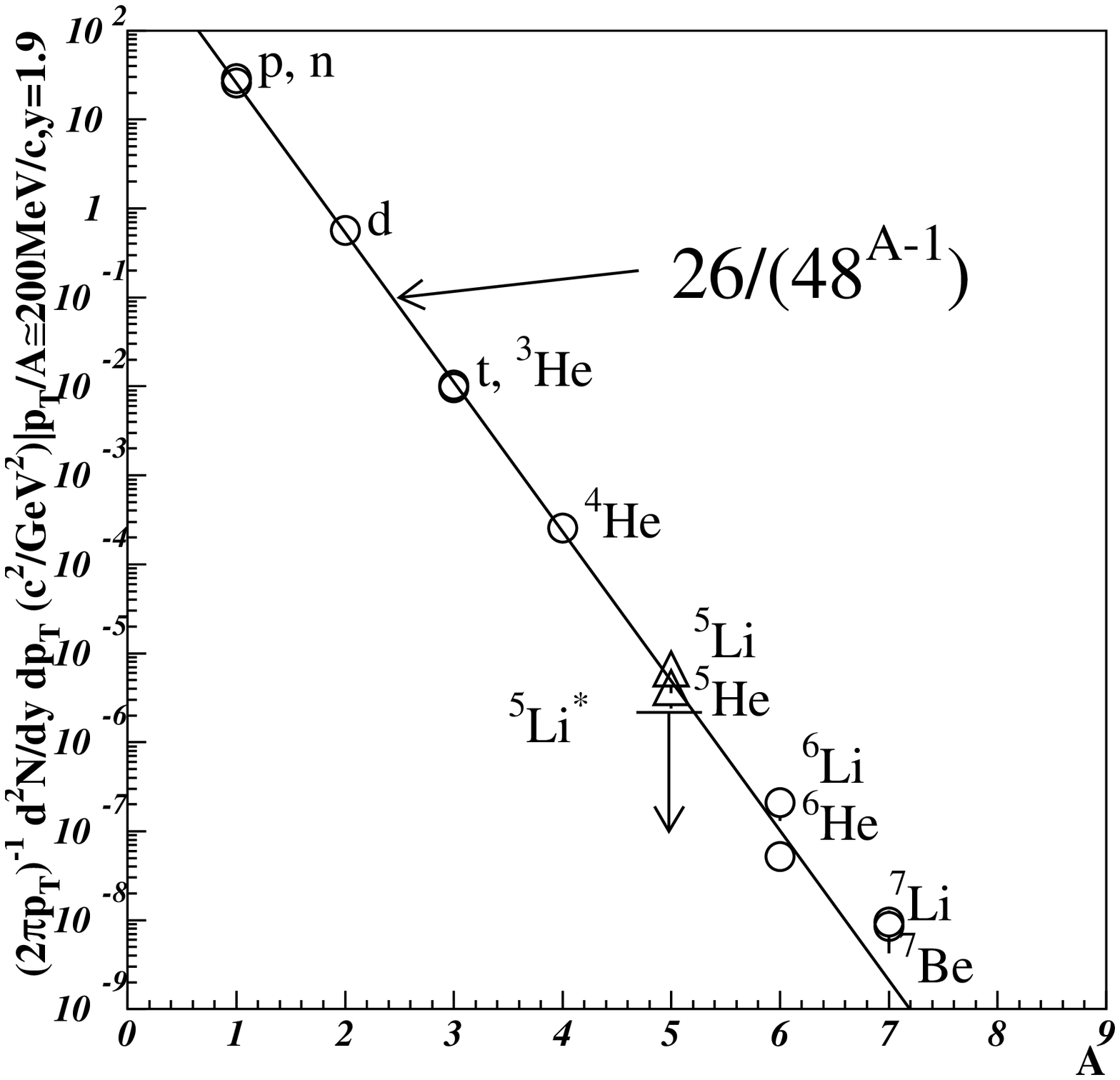}
\vspace{-0.45in}
\caption{Invariant yields at or near $y=1.9$, $p_{T}/A=200 MeV/c$ as a function of mass number $A$.
Strongly decaying states are represented by triangles and the $^{5}Li^{*}$ bar represents an upper limit.}
\label{fig:adep}
\end{minipage}
\end{figure}

In Figure~\ref{fig:temp_vs_mass} we display inverse slope parameters
for protons, deuterons, and $^{3}He$ nuclei as a function of mass number.
These slope parameters are determined from Boltzmann fits in transverse 
mass within the rapidity range from 2.2 to 2.4 and are shown both for 10\% 
most central events and more peripheral events.  Polleri et. al. \cite{Polleri}
have demonstrated that such trends in the temperature 
parameters of light nuclei are sensitive to both the spatial profile
of coalescing nucleons and to the flow profile of the source.  Our reach
in mass up to $A$=3 is then very useful in untangling these two profiles.

If we focus only on a narrow slice in transverse momentum
at $p_{T}/A$=100-200 MeV/c, we can examine the yields of these light
nuclei as function of rapidity.  We find that the invariant
yields are concave in shape (i.e. are lowest at midrapidity and
increase toward beam rapidity) and that the relative concavity increases
as we move to nuclei with higher masses.  This can be 
understood as an effect of longitudinal flow or incomplete stopping.

These and other trends in these lighter nuclei are explored much more
fully in \cite{nkg_parkcity}.
 
\vspace{-0.05in}
\section{HEAVIER STABLE LIGHT NUCLEI}

We now focus only on 10\% most central events
and on a kinematic region around $y$=1.9, $p_{T}/A$=200 MeV/c 
and explore the dependences of production on properties of the nuclei.

To start simply, we plot in Figure~\ref{fig:adep}
the invariant yields in this narrow kinematic region as a function of mass
number $A$ for $A \leq 7$.  We see that over almost ten orders of
magnitude, the yields are well described by a simple exponential
dependence; implying that to add a nucleon to a cluster, one
pays a penalty factor of approximately 48.  There are small deviations
from this trend, and we next move to trying to explain those.

Two dependences that are apparent are the spin and isospin dependences
of the yields.  Examining the ratios of invariant yields for $n/p$, $t/^{3}He$ 
and $^{6}He/^{6}Li$ near this same kinematic range, we find that $n/p$
and $t/^{3}He$ are consistent with a value of approximately 1.2 while
the $^{6}He/^{6}Li$ ratio is nearer to 0.3.  The neutron, proton,
triton, and $^{3}He$ are all spin $J$=1/2, while $^{6}He$ is $J$=0
and $^{6}Li$ is $J$=1.  We take this as evidence that the
yields of these nuclei include a dependence of $1.2^{-I_{Z}} \times (2J+1)$ 
(with $I_{Z}=Z-A/2$) where
the first term accounts for the neutron to proton ratio at freeze-out and
the second term is for the spin degeneracy of the state as is generally included
in coalescence models.

\begin{figure}[htb]
\begin{minipage}[b]{.46\linewidth}
\centering\epsfxsize=3in \epsfysize=3in \leavevmode \epsfbox[10 154 534 654]{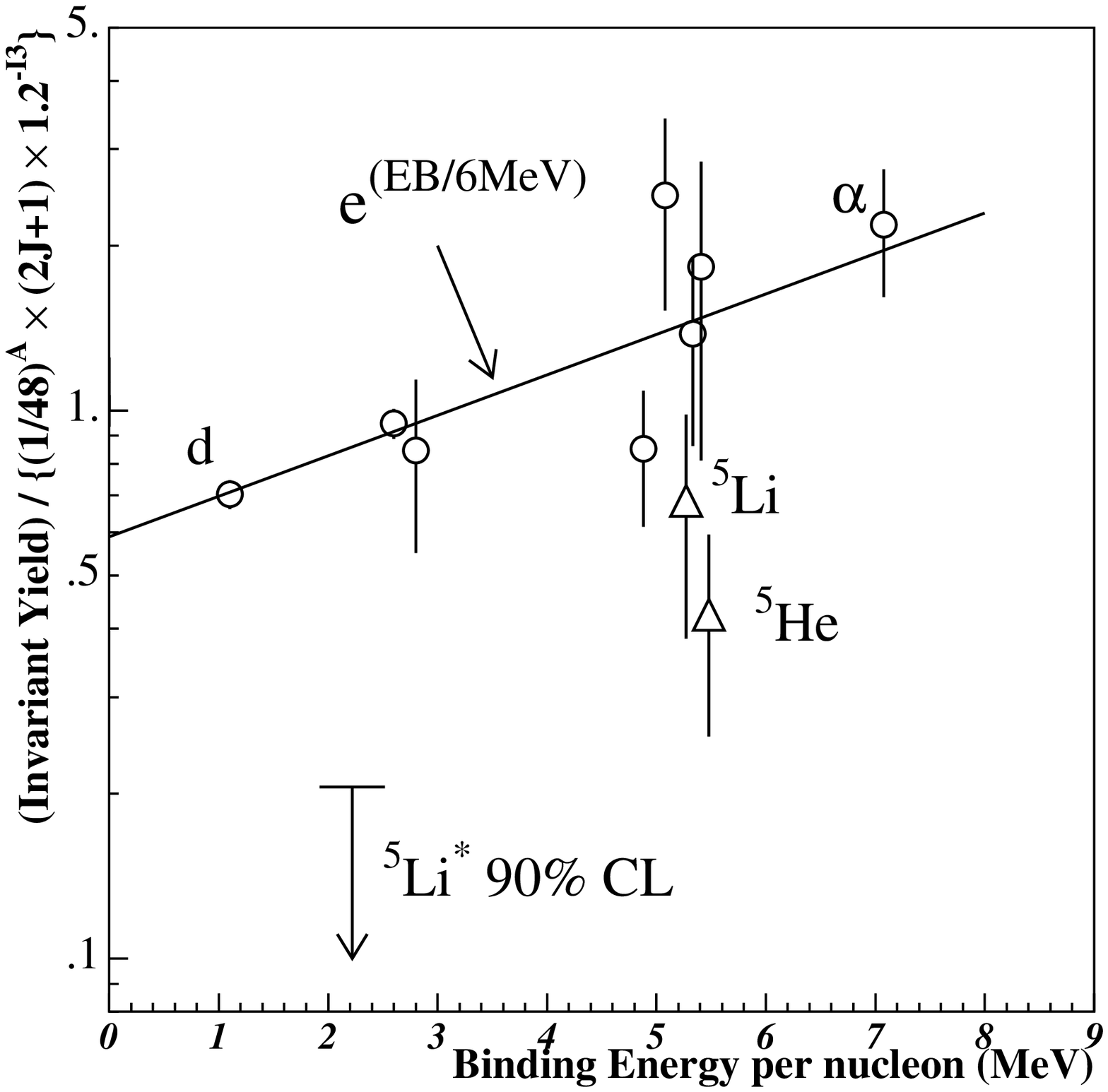}
\vspace{-0.45in}
\caption{Invariant yields of light nuclei with mass, spin, and isospin dependences divided out
as indicated, plotted versus binding energy per nucleon.  Strongly decaying states are denoted
by triangles.}
\label{fig:bind}
\end{minipage}
\hspace{\fill}
\begin{minipage}[b]{.46\linewidth}
\centering\epsfxsize=3in \epsfysize=3in \leavevmode \epsfbox[10 154 534 654]{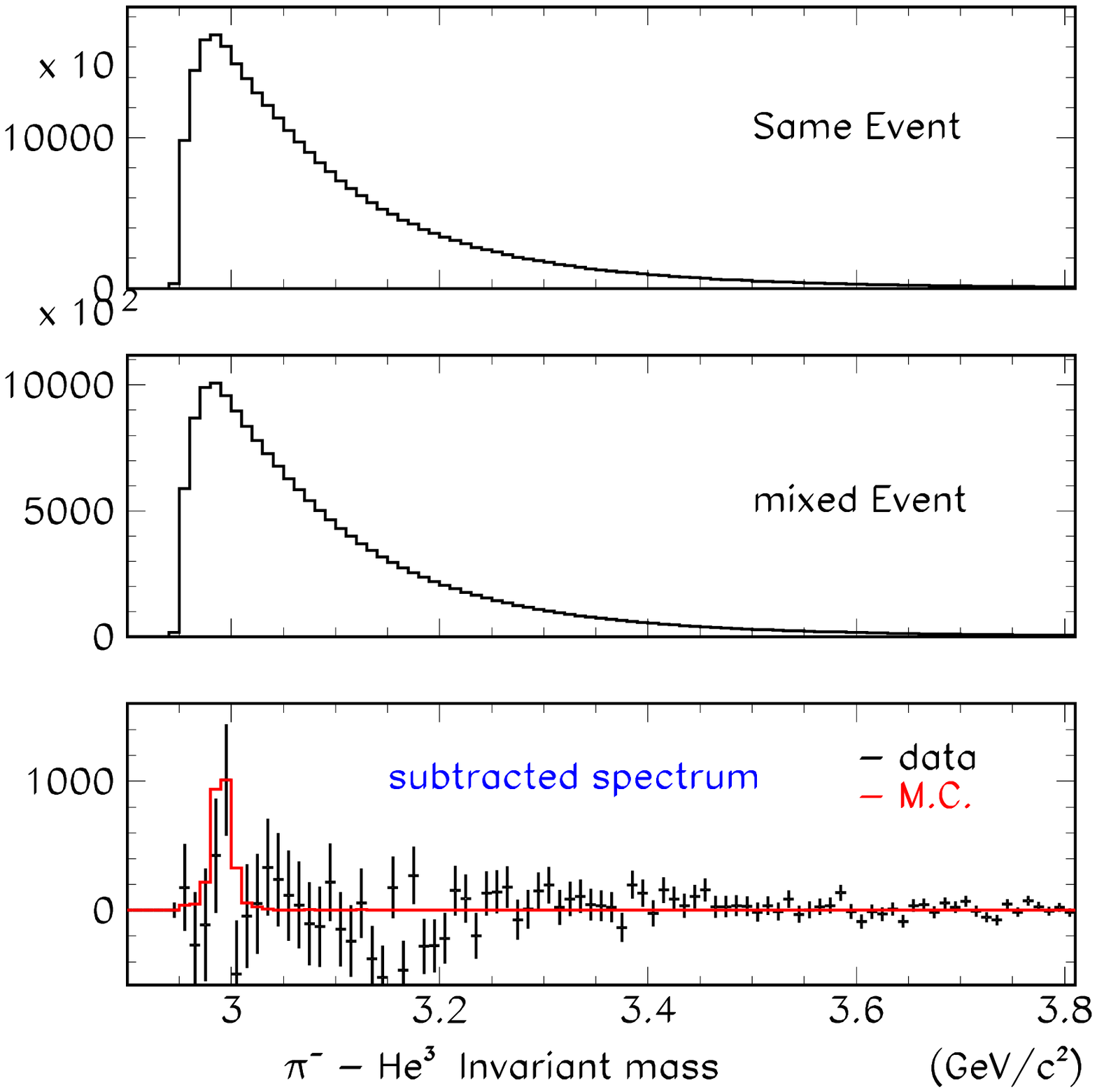}
\vspace{-0.45in}
\caption{Histograms of invariant mass of ($^{3}He,\pi ^{-}$) pairs showing pairs from
the same event (top), mixed events (middle), and 
the difference between the two with a monte-carlo prediction of the signal.}
\label{fig:h3l}
\end{minipage}
\end{figure}

We next take these measured invariant yields from Figure~\ref{fig:adep} and
divide them by $(1/48)^{A} \times 1.2^{-I_{Z}} \times (2J+1)$ to remove
these dependences on mass number, spin, and isospin of the states.  We then
look for any other trends in these 'corrected' yields.  In fact, we do see such a trend
when we plot these ratios as a function of binding energy per nucleon as shown
in Figure~\ref{fig:bind}.  This binding energy per nucleon
is defined as the sum of all the masses of the constituent nucleons of a state
less the nuclear mass all divided by the number of nucleons.  We note the
low temperature parameter of 6 MeV that we extract from this fit (strongly
decaying states were not included in determining the fit). 

It is possible then that this represents some final freeze-out of the nucleons
at a much lower temperature than kinetic freeze-out, or it is possible that
this is evidence of some other trend that coincidently shows up as a correlation
with this binding energy.  We are also exploring other possible correlations such as
with nuclear size.

\vspace{-0.20in}
\section{STRONGLY DECAYING STATES AND HYPERFRAGMENTS}

To complement these measurements of stable light nuclei, we have made measurements
of strongly decaying nuclei and the hypernucleus $^{3}_{\Lambda}H$.  
Because these analyses both involve states which decay at most centimeters from
our target (and therefore have decay products which are indistinguishable in our 
apparatus from particles originating from the target), they proceed along similar
lines.

The measurements are made in each case by identifying both of the decay products in our
apparatus.  When a pair of decay products is found in an event, an invariant
mass for the pair is calculated and placed in a histogram of pair invariant masses.
To mimic and subtract away the background, pairs consisting of one decay product
from one event and the second from a separate event are put though this same
process, producing a second invariant mass histogram.  The properly normalized 
difference of these two histograms should then be the signal under study as shown
in Figure~\ref{fig:h3l} (see \cite{rm} and \cite{sb} for details).

With this method we measure $^{5}Li$ ($^{5}He$) through its decay to $p(n)+ \alpha$.  
On a gross scale, the invariant yields
of these states sit closely along the trend shown in Figure~\ref{fig:adep}.
But on closer inspection in Figure~\ref{fig:bind}, the yields
for these states sit lower by a factor of 2 to 4 than the trend set by
the stable light nuclei.

Perhaps still more interesting is the absence of any signal for the $^{5}Li^{*}$ 
16.7 MeV excited state from its decay to $d + ^{3}He$.  From the absence
of a signal, we set the 90\% confidence level upper limit on production of this
state as shown in Figures~\ref{fig:adep} and~\ref{fig:bind}. Note that if we
assume that the ground state production sits at its most probable value from
our measurement, then this upper limit implies an upper limit for an 'excited state
temperature' of 15 MeV in a simple thermal picture; this is consistent with 
measurements \cite{gerd} made at the lower 
energies of GSI and possibly indicative \cite{gerd} of the same final-state 
interactions that may give rise to the the trend seen in Figure~\ref{fig:bind}.  


After the analysis of nearly 2/3 of the relevant data,
we observe a signal for $^{3}_{\Lambda}H$ at approximately the $2 \sigma$ level.
When our analysis is complete, we expect if this is indeed a signal that with improved analysis 
methods and the use of the full statistics we will observe a signal at approximately the $3 \sigma$ level.  

We measure a preliminary invariant multiplicity for $^{3}_{\Lambda}H$ in
the kinematic range $1.6<y<2.8, 0<p_{T}<2 GeV$ of $2.6 \pm 1.4 \times 10^{-4}$ $c^{2}/GeV^{2}$.
We can then extract a total penalty factor for adding a unit of 
strangeness in this kinematic range by taking the ratio of the $^{3}_{\Lambda}H$
invariant yield to that of $^{3}He$ in the same kinematic region.  This
gives a penalty factor of $Y(^{3}_{\Lambda}H)/Y(^{3}He) = $ $.031 \pm .018$.  Some
of this penalty is simply due to the fact that it is harder to 
get the all ingredients of a $^{3}_{\Lambda}H$ close enough in momentum space
to coalesce than is the case for $^{3}He$.  We can remove this part
of the penalty factor by normalizing each of these yields to the product
of the invariant yields of their ingredient baryons in the relevant kinematic 
ranges.  Using results from E891 
for the $\Lambda$ yields and E864 measurements
for the remaining quantities, we obtain 
$\frac{Y(^{3}_{\Lambda}H)}{Y(n) \times Y(p) \times Y(\lambda)} / \frac{Y(^{3}He)}{Y(n) \times Y^{2}(p)} =  
.162 \pm .088$ as a preliminary result.
This then represents a penalty of approximately 6 for coalescing a strange as opposed to a non-strange
state; we note that this was predicted to be equal to 1 \cite{dover}.

\vspace{-0.02in}
\section{SUMMARY AND ACKNOWLEDGEMENTS}

By our measurements of light nuclei with $A \leq 3$ we can study trends in 
light nuclei production as a function of kinematics and centrality.  
In 10\% central collisions in a limited kinematic range, 
we can explore dependences of production on mass, spin, isospin, strangeness, and 
potentially binding energy of the state up to mass number $A = 7$.

An interestingly low 'excited state temperature' of less than 15 MeV is implied by our upper limit
on production of $^{5}Li^{*}$, and our measurement of $^{3}_{\Lambda}H$
implies a larger than expected strangeness penalty factor.

We gratefully acknowledge the efforts of the AGS staff and support
from the DOE High Energy and Nuclear Physics Divisions and the NSF.


\end{document}